\documentclass[12pt,preprint]{aastex}

\usepackage{CJK}

\usepackage{epsfig}
\usepackage{amsmath}

\shorttitle{\ion{H}{1} FREE-BOUND EMISSION of PNe}
\shortauthors{Zhang \& Liu}

\begin{document}

\title{{\ion{H}{1}} FREE-BOUND EMISSION OF PLANETARY NEBULAE WITH LARGE
ABUNDANCE DISCREPANCIES: TWO-COMPONENT MODELS VERSUS $\kappa$-DISTRIBUTED ELECTRONS
}
\begin{CJK*}{UTF8}{gbsn}

\author{Yong Zhang (张泳)$^{1}$, Xiao-Wei Liu (刘晓为)$^{2,3}$, and Bing Zhang (张兵)$^2$}

 \altaffiltext{1}{Department of Physics, The University of Hong Kong, Pokfulam Road, Hong Kong, China; zhangy96@hku.hk}
\altaffiltext{2}{Department of Astronomy, Peking University, Beijing 100871, China}
\altaffiltext{3}{Kavli Institute for Astronomy and Astrophysics, Peking University, Beijing 100871, China}

\begin{abstract}

The ``abundance discrepancy'' problem in the study of planetary nebulae (PNe),
viz., the problem concerning systematically higher heavy-element abundances
derived from optical recombination lines relative to those from 
collisionally excited lines, has been under discussion for decades,
 but no consensus on its solution has yet been reached. 
In this paper we investigate the hydrogen free-bound emission near the Balmer jump 
region of four PNe that are among those with the largest abundance discrepancies,
aiming to examine two recently proposed solutions to this problem: 
two-component models and $\kappa$ electron energy distributions.
We find that the Balmer jump intensities and the spectrum slopes
cannot be simultaneously matched
by the theoretical calculations based upon single Maxwell-Boltzmann 
electron-energy distributions, whereas the fitting can be
equally improved by introducing $\kappa$ electron energy distributions
or an additional Maxwell-Boltzmann component.
We show that although \ion{H}{1} free-bound emission alone cannot
distinguish the two scenarios, it can provide important
constraints on the electron energy distributions, 
especially for cold and low-$\kappa$ plasmas.

\end{abstract}

\keywords{atomic processes --- ISM: abundances --- planetary nebulae: general
--- plasmas
}

\maketitle
\end{CJK*}

\section{INTRODUCTION}


The determination of element abundances in planetary nebulae (PNe) is 
essential for understanding the stellar nucleosynthesis processes and the chemical enrichment in the interstellar medium. However, a number of studies of PNe
have consistently established an intriguing puzzle that heavy-element abundances
determined from optical recombination lines (ORLs) are systematically higher than 
those derived from collisionally excited lines (CELs).
The abundance discrepancy is commonly quantified by the ratio between the 
O$^{2+}$ abundances obtained from ORLs and CELs, called abundance discrepancy
factor (ADF). The CEL abundance determination is subject to an accurate determination 
of the electron temperature. A relevant problem is that the electron 
temperatures 
obtained from the \ion{H}{1} Balmer jump are  generally lower than  those 
obtained from [\ion{O}{3}] CELs (to be referred to as $T_{\rm e}$(BJ) and $T_{\rm e}$([\ion{O}{3}]), respectively, hereafter).  The ADFs have been found to be positively
correlated with the temperature differences \citep{ll01}.
Several explanations for the abundance and temperature discrepancies have been proposed, but no consensus has 
emerged \citep[see,][and the references therein for further details on the two problems]{pei67,liu93,st04,liu06,pei06}, among which we will focus on two 
here: two-component models and $\kappa$-distributed electrons.

The two-component models  assume that there exist spatially 
unresolved knots within diffuse nebulae \citep{ls00}. These knots 
are extremely metal-rich and partially or fully ionized.
Because of high cooling rates, they have relatively low temperatures,
which therefore greatly favor the emission of ORLs and suppress that 
of CELs. It follows that in the scenario of two-component models, ORLs
and CELs inform us about the abundances and electron temperatures in different 
nebular components. Detailed three-dimensional photoionization models of
NGC\,6153, a PN exhibiting a large ADF, show that the two-component models
incorporating a small amount of metal-rich inclusions can successfully reproduce
all the observations \citep{yuan11}. The chemical pattern can exclude the ejection of 
stellar nucleosynthesis products as the origin of these knots.
\citet{liu06} suggested that they  might be produced by evaporating
planetesimals. The hypothesis of solid body destruction was theoretically investigated 
by \citet{hen10}, who concluded that under certain conditions
the sublimation of volatile bodies possibly produces
enough metal-rich gas to explain the ORL/CEL abundance discrepancies.
The main criticism of this model is that there is no direct observational
evidence for the existance of such knots.

Previous calculations of  element abundances and electron temperatures
in PNe have been based upon a widely accepted assumption that free electrons 
have Maxwell-Boltzmann (M-B) energy distributions \citep{spi48}. Recently, \citet{nic12,nic13} 
presented that the presence of non-thermal electrons can potentially account for
the abundance and temperature discrepancy problems. In this scenario, the free 
electrons have non-equilibrium energy distributions whose departure from
M-B distributions can be parameterized by a $\kappa$ index. Such a $\kappa$
distribution involves a low-temperature M-B core and a power-law high energy 
tail, and has been applied to fit the energy distributions of energetic particles
in solar system plasmas \citep[e.g.][]{liv11a}.
Since the low- and high-energy electrons are preferential for recombination and
collision processes, respectively, ORLs would indicate a lower electron 
temperature than CELs if using an M-B distribution to interpret the spectrum 
with a $\kappa$ distribution.
\citet{leu02} and \citet{liv09} show that the phenomenologically 
introduced $\kappa$ distributions in the studies of solar system plasmas
can arise naturally
from Tsallis' nonextensive statistical mechanics\footnote{Tsallis' 
nonextensive statistical mechanics statistics is a generalization of the 
conventional Boltzmann-Gibbs Statistics, in which an entropic index
$q$ is introduced to characterize the degree of non-additivity of the
system. The $q$ and $\kappa$ indexes are related with each other
through the simple equation $q=1+1/\kappa$.}. 
In collisionless plasmas, it takes a long time for high-energy electrons to 
relax to their equilibrium distribution, which thus can reach 
a non-equilibrium stationary state if they can be continuously pumped.
The physical mechanisms producing 
non-thermal electrons in PNe, however, have never been thoroughly investigated, although \citet{nic13} sketched a few possibilities. 

Apparently, measuring the electron energy distribution from 
observations is important to examine the two propositions. For this purpose,
\citet{ss13} studied \ion{C}{2} dielectronic recombination lines in a few PNe, but the uncertainties are too large to draw any definite conclusion.
To further investigate the two scenarios, in this paper we 
study the hydrogen free-bound continua of four PNe with very large ADFs ($>6$).
In Section~2, we  present the methodology and discuss the possibility of using 
hydrogen free-bound continua as a probe of the electron energy distribution. 
In Section~3, we fit the observed spectra with theoretical calculations of
two-component models and $\kappa$-distributed electrons, and discuss the
implications of  these results. Section~4 is a summary of our conclusions.

\section{METHODOLOGY}

The \ion{H}{1} free-bound spectrum is emitted when a free electron is captured by a 
proton, and thus can potentially sample the energy distribution of recombining electrons.
Below we present the calculations of \ion{H}{1} free-bound spectra near the Balmer
jump region in M-B and $\kappa$ electron energy distributions, followed by
 a brief description of  the PN sample and the spectral fitting procedure.

\subsection{\ion{H}{1} free-bound emission from M-B electron distributions\label{fbemissin}}

The \ion{H}{1} continuous spectrum is calculated  using the same method as described in 
\citet{zhang04}. Assuming that free electrons have an M-B energy distribution, viz.,
\begin{equation}\label{MB1}
\frac{n(E)}{N}dE=2\sqrt{\frac{E}{\pi}}\left(\frac{1}{k_BT_U}\right)^{3/2}{\rm e}^{-E/k_BT_U}dE,
\end{equation}
 the emission coefficient of \ion{H}{1} free-bound 
continuous emission is given by
\begin{equation}
 j_{\nu}^{\rm M-B}=\frac{1}{4\pi}N_{\rm p}N_{\rm e}\frac{\pi h^4\nu^3}{c^2}\left(\frac{2}{\pi m_{\rm e}k_BT_U}\right)^{3/2}\sum_{n=n_1}^{\infty}\sum_{l=0}^{n-1}n^2a_{nl}(\nu){\rm e}^{h(\nu _{nl}-\nu)/k_BT_U}, \label{MB2}
\end{equation}
where $k_B$ and $h$ are the Boltzmann and the Planck constants, respectively, 
$c$ is the speed of light, $m_{\rm e}$ is the electron mass,
$N_{\rm p}$ and $N_{\rm e}$ are proton and electron densities respectively, 
$h\nu _{nl}$ is the ionization potential of the ($n$,$l$) state, $n_1$
is the lowest state that can contribute to the free-bound emission at the
given frequency $\nu$  ($n_1=2$ and 3 for the Balmer and Paschen recombination continua,
respectively), and $a_{nl}(\nu)$ is the photoionization cross-section
for the state, which is computed using the method described by \citet{storey1991}.
$T_U$ is the kinetic temperature reflecting the internal energy of free
electrons, which under the M-B distribution 
is commonly referred to as electron temperature\footnote{$T_{\rm e}$ represents electron equilibrium temperature. Under the $\kappa$ distribution,
$T_U$ is non-equilibrium temperature.}, $T_{\rm e}$. 
Free-free emission and two-photon decay are also taken into account 
\citep[see][for the details]{zhang04}. The spectrum then is normalized 
to the integrated intensity of the H11 Balmer line at 3770\,{\AA}, whose emission coefficient is given by
\begin{equation}
 j_{{\rm H}11}^{\rm M-B}=\frac{1}{4\pi}N_{\rm p}N_{\rm e}\alpha^{\rm eff}_{{\rm H}11}h\nu_{{\rm H}11},
\end{equation}
where $\alpha^{\rm eff}_{{\rm H}11}$ is the effective recombination coefficient of
the H11 line derived by \citet{hummer1987} based upon M-B electron distributions.
Hereafter, we use $I_\lambda$, in units of {\AA$^{-1}$}, to denote the normalized continuum
intensity.

In Figure~\ref{MB} we show the calculated \ion{H}{1} free-bound spectra near the
Balmer jump region at different $T_U$ values along with the corresponding electron 
energy distributions.  It is apparent from this figure that the shapes 
of theoretical spectra 
strongly depend on the electron energy distributions. We choose four positions, as
labeled in Figure~\ref{MB} (`abcd', where $\lambda_a=3500$\,{\AA}, 
$\lambda_b=3650^-$\,{\AA, $\lambda_c=3650^+$\,{\AA, and $\lambda_d=3900$\,{\AA}), 
to characterize the spectra, and thus the spectral shapes can be quantitatively
described with the slopes of the Balmer continuum $S_B=(I_b-I_a)/(\lambda_a-\lambda_b)$ and 
the Balmer jump $J_B=I_b-I_c$. 
The a--b and c--d spectra are respectively produced by free electrons recombining to
$n=2$ and 3 states, whose kinetic energies satisfy the relation $E_b(0\,{\rm eV})<E_a(0.14\,{\rm eV})<E_d(1.67\,{\rm eV})<E_c(1.89\,{\rm eV})$.
An inspection of Figure~\ref{MB} reveals that with increasing kinetic temperatures, 
$S_B$ and $J_B$ decrease while $I_c$ and $I_d$ increase. This is clearly due to
the increasing number of high-energy electrons with respect to low-energy electrons.
At high kinetic temperatures ($T_U>10000$\,K), the $S_B$ and $J_B$ values become 
significantly less sensitive to $T_U$, attributing to the incapability of
\ion{H}{1} continua near the Balmer jump region to trace very energetic electrons.
Consequently, the examination of Figure~\ref{MB} indicates that
\ion{H}{1} free-bound continua can provide a potential probe of
electron energy distributions in cold plasma.

\subsection{\ion{H}{1} free-bound emission from $\kappa$ electron distributions}

The electron energy distributions in the vicinity of the Sun and a few planets
have been found to have an M-B core and  an enhanced high-energy tail, which
can be well described by a $\kappa$ distribution \citep[i.e.][]{liv09,liv11} defined by
\begin{equation}\label{k1}
\frac{n(E)}{N}dE=2\sqrt{\frac{E}{\pi}}\left(\frac{1}{k_BT_U}\right)^{3/2}
\frac{\Gamma(\kappa+1)}{(\kappa-3/2)^{3/2}\Gamma(\kappa-1/2)}\left[1+\frac{h(\nu_{nl}-\nu)}{(\kappa-3/2)k_BT_U}\right]^{-\kappa-1},
\end{equation}
where $\Gamma$ is the gamma function and $\kappa$ is a parameter of $>1.5$ 
describing the degree of departure from the M-B distribution.
This inspired \citet{nic12} to suggest that the $\kappa$ distribution also 
applies for the plasmas in PNe. 
As illustrated in \citet{nic12}, at a given $T_U$, 
decreasing $\kappa$ values would cause a shift of the number of 
intermediate-energy electrons towards those of
higher and lower energies. As a result, the
low-energy region manifests its profile as an M-B distribution with an equilibrium 
temperature\footnote{Namely, 
the $n(E)/N$ values of a $\kappa$ distribution at low-energy region
can match those of an M-B distribution with a temperature $T_{\rm core}$
by scaling a factor of $>1$ \citep[see][]{nic13}.}
\begin{equation}\label{tt}
T_{\rm core}=\left(1-\frac{3}{2\kappa}\right)T_U.
\end{equation}

In the case of the $\kappa$ electron energy distribution,
the emission coefficient of \ion{H}{1} free-bound continuous emission can be written as
\begin{equation}\label{k2}
\begin{split}
 j_{\nu}^{\kappa}=&\frac{1}{4\pi}N_{\rm p}N_{\rm e}\frac{\pi h^4\nu^3}{c^2}\left(\frac{2}{\pi m_{\rm e}k_BT_U}\right)^{3/2}\frac{\Gamma(\kappa+1)}{(\kappa-3/2)^{3/2}\Gamma(\kappa-1/2)}\\
&\times\sum_{n=n_1}^{\infty}\sum_{l=0}^{n-1}n^2a_{nl}(\nu)
\left[1+\frac{h(\nu_{nl}-\nu)}{(\kappa-3/2)k_BT_U}\right]^{-\kappa-1}.
\end{split}
\end{equation}
In the limit $\kappa\rightarrow\infty$, the $\kappa$ distribution tends to
the M-B, and Equations~(\ref{k1}) and (\ref{k2}) go over to 
Equations~(\ref{MB1}) and (\ref{MB2}), respectively.


We investigated the behavior of \ion{H}{1} free-bound spectra under
$\kappa$ electron distributions. Figure~\ref{kappa} displays the calculated spectra 
at a given temperature $T_U=10000$\,K but different $\kappa$ values, as well as the
 corresponding electron energy distributions. In order to obtain the normalized spectra,
the recombination coefficient of the H11 line has been corrected according to
Equation~(17) in \citet{nic13}, so that the emission coefficient is
given by
\begin{equation}
j_{{\rm H}11}^\kappa=\frac{(1-3/2\kappa)\Gamma(\kappa+1)}{(\kappa-3/2)^{3/2}\Gamma(\kappa-1/2)}j_{{\rm H}11}^{\rm M-B}.
\end{equation}
 As shown in Figure~\ref{kappa}, decreasing
$\kappa$ values cause the peak of the electron distribution to move to lower energies
(see the lower panel),
resulting in increasing $J_B$ and $S_B$ and decreasing $I_c$ and $I_d$
(see the upper panel).
The theoretical spectrum at a high $\kappa$ value is almost identical to that
calculated based upon an M-B electron distribution of the same
temperature. Therefore, in principle,
one can determine the $\kappa$ value by comparing the theoretical and observed
 \ion{H}{1} free-bound continua. This method is particularly suitable
when the $\kappa$ value is low and thus the distribution of low-energy electrons
is sensitive to its alteration.  Because \ion{H}{1} free-bound continua are 
insensitive to the $\kappa$ value when $\kappa$ is large, we choose a few
extreme PNe with large ADFs for this study, which are supposed to have extremely
low $\kappa$ values if the abundance discrepancies are caused by
the $\kappa$ electron distribution.


In order to address the question whether the \ion{H}{1} free-bound spectrum allows one to distinguish between $\kappa$ and M-B electron distributions,
in Figure~\ref{comp} we compared the spectra calculated based upon a $\kappa$ distribution at $T_U$ and an M-B distribution at $T_{\rm core}$, where $T_U$ and $T_{\rm core}$ satisfy the 
relation given by Equation~(\ref{tt}). As shown in this figure, the Balmer
continua (a--b) of the 10000\,K $\kappa$ distribution and the 2500\,K M-B
distribution are nearly parallel, viz., the $S_B$ values are equal, which can be
attributed to the fact that the profile of the cold M-B electron distribution 
closely resembles that of the low-energy region of the $\kappa$ one.
However, the 2500\,K M-B distribution predicts a larger $J_B$ value, which
is conceivable since its peak $n(E)/E$ value is larger (see the lower panel of
Figure~\ref{comp}). Since the temperature-only-dependent M-B distributions indicate a
one-to-one correspondence between $S_B$ and $J_B$ (Figure~\ref{MB}), it follows
that the \ion{H}{1} free-bound continua provide a diagnostic to
separate the two kinds of electron distributions,
and $S_B$ and $J_B$ can be used to evaluate the $\kappa$ value.
Another implication of our calculations is that using an M-B electron 
distribution
one is unable to simultaneously match the $S_B$ and $J_B$ values
of the \ion{H}{1} free-bound continua arising from a $\kappa$ electron distribution, in that $J_B$ will imply a higher M-B temperature than $S_B$.

\subsection{The sample and spectral fitting}

Our sample includes four PNe, Hf\,2-2, NGC\,6153, M\,1-42, and M\,2-36,
which are among the PNe exhibiting the largest temperature and abundance
discrepancies. The high signal-to-noise spectra are taken from 
\citet{ls00,ll01,lb06}, whose primary purpose was to
 measure the intensities of weak ORLs from heavy-element ions.
Careful treatments have been devoted to flux calibrations and dereddening
corrections \citep[see][for the details]{ls00,ll01,lb06}.
Table~\ref{para} gives the ADFs, $T_{\rm e}$(BJ), and $T_{\rm e}$([\ion{O}{3}])
deduced through the empirical analysis.
Their ADFs range from 6.9 to 70,  much larger than the average value of
Galactic PNe ($\sim2$).
The $T_{\rm e}$([\ion{O}{3}]) values are found to be higher than $T_{\rm e}$(BJ)
by a factor of 1.4--9.4.

In order to fit the observed \ion{H}{1} free-bound continua, we consider
three possibilities for electron energy distributions:  single M-B 
distributions,  bi-M-B distributions, and $\kappa$ distributions. 
The former has a single fitting parameter (i.e. the electron temperature), while the later two have 
two parameters (see the next section). The reduced
$\chi^2$ values were calculated
in a parameter space for each object to evaluate the goodness of fit, 
which is defined by 
\begin{equation}
\chi^2=\frac{1}{\eta}\sum \frac{(I_{\lambda, \rm obs}-I_{\lambda, \rm the})^2}{\sigma_\lambda^2},
\end{equation}
where $I_{\lambda, \rm obs}$ and $I_{\lambda, \rm the}$ are the observational 
and 
theoretical continuum intensities in the line-free regions of the spectrum 
ranging from 3200--4200\,{\AA}, $\sigma_\lambda$ is the measurement error of
$I_{\lambda, \rm obs}$ caused by noises, and $\eta$ is the  number of degrees of freedom.
The optimal fitting for each object is then achieved with the temperature and/or other fitting parameters that yield the minimum $\chi^2$ value.

In addition to \ion{H}{1} recombination continua, the observed spectra
contain a contamination from the direct or scattered light from the central 
star.  
  Consequently, the theoretical continuum intensity is given by
\begin{equation}
I_{\lambda,\rm the}= I_{\lambda,\rm star}+I_{\lambda,\rm H},
\end{equation}
where the theoretical \ion{H}{1} recombination intensity
$I_{\lambda,\rm H}$ is a sum of contributions from
free-bound, free-free, and two-photon emission, among which 
the free-bound transition dominates the spectrum near the
Balmer jump, and the free-free emission is negligible.
In order to minimize the number of fitting parameters, we assume that the spectral energy distribution of
the contaminating stellar continuum follows a power law,
\begin{equation}
I_{\lambda,\rm star} = (I_{b, \rm obs}-I_{b,\rm H})(\lambda_b/\lambda)^{-\beta},
\end{equation}
where $I_{b, \rm obs}$ and $I_{b, \rm H}$ are the observational 
continuum intensity and theoretical
\ion{H}{1} recombination
intensity at $\lambda_b=3650^-$\,{\AA}, respectively.  
$I_{\lambda,\rm star}$ is initially assumed
to follow a Rayleigh-Jeans approximation to a blackbody, viz., $\beta=4$.
However, the actual situation might be more complex because of
the wavelength and spatial dependence of scattered light. Moreover, there
is some uncertainties in the emission coefficient of the two-photon process
that can affect the calculations of continuum intensity
\citep[see][]{zhang04}. Therefore, $\beta$ is slightly adjustable, and
we adopt $\beta$ to be an integer between 2 and 5 yielding the best fit.

\section{RESULTS and DISCUSSION}

The fitting results are shown in Table~\ref{para}. We find that for all
the PNe the observed continua cannot be successfully fitted with the
theoretical spectra calculated by Equation~\ref{MB2} for
 single M-B electron energy distributions. In Figure~\ref{one} we plot
the temperature dependences of $\chi^2$  in the one-parameter model.
The distributions of filled circles in Figure~\ref{one} are less dense in the low-temperature side, attributing to the fact that the sensitivity of
\ion{H}{1} recombination spectra to temperature increases
with decreasing $T_{\rm e}$.  The minimum $\chi^2$ values 
in the range of $300\,{\rm K}<T_{\rm e}<25000\,{\rm K}$ are 
unacceptably large ($>4$).
Specifically, $S_B$ and $J_B$ cannot be 
simultaneously matched by the one-parameter model. As indicated
in Figure~\ref{fail}, the temperatures determined by $J_B$ are higher
than those by $S_B$. This is a clear evidence that there is 
an excess of high-energy electrons with respect to the M-B 
distribution indicated by low-energy electrons 
(i.e. the fractional population of  electrons at $E_d$ and $E_c$
is higher than that inferred from the electrons at $E_a$ and $E_b$; 
see Figure~\ref{MB}). Qualitatively saying, the excess of high-energy electrons
can be caused by either bi-B-M (the two-component model) or $\kappa$ 
electron energy distributions. Below we investigate the two scenarios.


The two-component model assumes that nebular spectra arise from two gaseous
components with different equilibrium temperatures, $T_h$ and $T_c$,
representing electron temperatures  of the hot diffuse nebulosities and
the cold metal-rich knots, respectively (hereafter super- or subscript ``$h$'' 
and ``$c$'' refer to quantities of the hot and cold compoents, respectively).  In this scenario, the theoretical
\ion{H}{1} recombination spectrum is a sum of contributions from the
two components with individual M-B electron distributions, with the 
intensity of 
\begin{equation}
I_{\lambda,\rm H} = \frac{f}{f+1}I_{\lambda,{\rm H}}^c+\frac{1}{f+1}I_{\lambda,{\rm H}}^h,
\end{equation}
where $f$ is the H11 line intensity ratio of the cold component
over the hot component.
 Thus we have
\begin{equation}
f=\frac{N_{{\rm p},c}M_{{\rm e},c}\alpha^{\rm eff}_{{\rm H}11,c}}{N_{{\rm p},h}M_{{\rm e},h}\alpha^{\rm eff}_{{\rm H}11,h}}, 
\end{equation}
where $M_{\rm e}$ is the total electron number of each component.
A detailed photoionization model
involving the two components has been successful in accounting for the spectrum
of NGC\,6153 \citep{yuan11}. Guided by the model of \citet{yuan11}, 
we assume that $T_h/T_c\equiv10$ and $N_{{\rm p},c}/N_{{\rm p},h}\equiv5$
for all the PNe. 
Consequently, two parameters, $T_h$ and $M_{{\rm e},c}/M_{{\rm e},h}$,
 are employed for the two-component model fitting. 

Through introducing an additional component, the fitting is greatly
improved (Table~\ref{para}). The best fits are shown in Figure~\ref{fitting}.
Figure~\ref{contourtwo} shows the $\chi^2$ distributions in the parameter
space that we search for. We find that this model is capable of
producing $S_B$ and $J_B$ by including only small amount of cold components
 with $M_{{\rm e},c}/M_{{\rm e},h}$ ratios of $<0.03$. 
The contours of the $\chi^2$  values elongate along the increment direction
of $T_h$ and $M_{{\rm e},c}/M_{{\rm e},h}$ (Figure~\ref{contourtwo}),
along which the fits are less sensitive to the choice of the fitting parameters.
This simply reflects the fact that to account for the observations 
the increasing inclusion of cold component will result in higher $T_h$
values. Through reconciling $T_h$ and $T_{\rm e}$([\ion{O}{3}]), the
temperature discrepancy problem can be solved.



In the case of $\kappa$ electron distributions, $I_{\lambda,\rm H}$
is calculated based upon Equation~(\ref{k2}), and $T_U$ and $\kappa$  are 
used as input parameters in the fitting procedure. Similar to the
two-component model, the introduction of $\kappa$ electron distributions
can significantly improve the fitting such that the minimum $\chi^2$ 
values can reach $<1.5$ (Table~\ref{para}). The $\chi^2$ distributions 
in the $T_U$-$\kappa$ space are shown in Figure~\ref{contourk}.
A decreasing $\kappa$ value can lead to an increment of $T_U$
to accommodate the observed $S_B$ and $J_B$.
The best fits, as shown in Figure~\ref{fitting}, indicate very low $\kappa$ 
values for all the PNe, suggesting a large departure from the M-B distribution 
for these high-ADF PNe.

The elongation direction  of the contours of $\chi^2$ 
(Figure~\ref{contourk}) is a reflection of the
fitting uncertainties, which suggests that  $\kappa$  can
be more accurately determined in the plasma  more significantly departing
from the M-B electron distribution. However,
there is a long tail toward large $\kappa$ values.
For NGC\,6153 and M\,1-42,  a $\kappa$ value of $\sim8$
can still achieve a reasonable match within the confidence level of 
$\chi^2<2.0$. Despite the large uncertainty, we can conclude that
the $\kappa$ value for Hf\,2-2 is extremely low. Such low $\kappa$ indexes have
been found in inner heliosheath, blazar $\gamma$-rays, solar flares, 
interplanetary shocks, corotating interaction regions, and solar wind
\citep{liv11}. Various theories have been developed to explain 
the $\kappa$ distributions in solar system plasmas \citep[][and the references therein]{pie10}.
The environments of PNe are very different. What physical mechanism
is responsible for producing such a large population of non-thermal 
electrons in PNe remains a subject for speculation.


We have shown that both the two-component models and $\kappa$ electron
distributions can provide better agreement with the observations
of \ion{H}{1} continua than  single M-B electron distributions.  
In the framework of a certain model,
the \ion{H}{1} continua behave differently for different parameter
settings, and thus can provide a diagnostic to determine these parameters.
In Figures~\ref{diabi} and \ref{dia}, we plot the plasma diagnostics
for the two-component models and $\kappa$ electron distributions,
respectively. An inspection of the two figures suggests that
 with accurate subtraction of stellar light contamination
it is possible to use $J_B$ and $S_B$ to derive
$T_U$, $M_{{\rm e},c}/M_{{\rm e},h}$, and $\kappa$.
The method is particularly useful for cold and/or low-$\kappa$
PNe.  
All the four PNe exhibit departure from the predictions of 
single M-B distributions (which coincide with those for
$M_{{\rm e},c}/M_{{\rm e},h}=0.0002$ and $\kappa=50$ respectively
in Figures~\ref{diabi} and \ref{dia}), the most extreme one of
which is Hf\,2-2.

In this study, we do not take into account high order hydrogen
recombination lines, which converge at the Balmer discontinuity (Figure~\ref{fail})
and are usually used to measure the electron density \citep[e.g.][]{zhang04}.
It is valuable to investigate their behavior under $\kappa$
electron distributions in the future.

The two-component models and $\kappa$-distributed electrons can
improve the fit of nebular spectra near the Balmer jump region
to a similar extent. Therefore, based on
the \ion{H}{1} free-bound continua alone, we cannot ascertain
which scenario is more appropriate. 
However, the two-component models are perhaps more physically motivated.
It suggests that PNe are spatially inhomogeneous in temperature, density,
and chemical abundance.  The cold knots embedded in diffuse nebulae are produced by the
destruction of solid bodies, leading to a contamination
of nebular spectra. Since the destruction of solid bodies
is related to their spatial distributions and evolutionary stages
of central stars, the two-component models can account for the
observations that ADFs are larger in more evolved PNe
and/or in the inner regions of a given PN. Furthermore, this model can
provide plausible explanations for the observational facts
that abundances derived from the temperature-insensitive infrared CELs
are comparable with those from optical CELs, and the ORL abundances
of high-ADF PNe are far beyond the predictions of stellar nucleosynthesis.
The $\kappa$ distributions have been familiar to
the solar physics community for decades. Such distributions have been proven
to better match the observations of solar system plasmas than bi-M-B
distributions. The $\kappa$ function was initially introduced
as a mathematical description of energy distributions.
As shown by \citet{leu02} and \citet{liv09}, a $\kappa$ distribution is a consequence of  
Tsallis's nonextensive statistical mechanics, yet its exact origin in solar system plasmas, 
and whether such an origin also works for photoionized gaseous nebulae, remains unclear.


\section{CONCLUSIONS}

In this paper, we investigate the \ion{H}{1} free-bound continua of four
PNe exhibiting extremely large ADFs. We find strong evidence that
these spectra are not emitted from the plasma with single M-B
electron energy distributions. Two possible explanations 
are examined: two-component models and $\kappa$-distributed electrons.
Our results show that both can adequately account for the observations.
We also present a method to determine the physical conditions
of PNe with two components or $\kappa$-distributed electrons.  
Specifically, the \ion{H}{1} Balmer jump $J_B$ and the slope
of Balmer continuum $S_B$ can be used to derive temperature,
filling factor of cold knots, and $\kappa$ index.
The results presented in this paper provide new insights
into the long-standing problem of abundance discrepancies in PNe.

Energy distributions of free electrons have a profound effect on abundance
calculations of PNe.  The recombination coefficients of ORLs and the collision
strengths of CELs rely on the integral of the known
energy dependence of the atomic cross section over the assumed distribution function.
Unlike for solar system plasmas, whose energy distributions can been accurately detected 
through the direct interaction between the plasma particles and the
detectors on-board satellites and space probes, it is rather
hard to measure the electron energy distributions of distant PNe.   
\ion{H}{1} free-bound spectra, which directly sample free electrons,
provide a promising approach for such a purpose, and
this paper can be regarded as a preliminary attempt.
In future studies, we will further investigate this problem by
obtaining high signal-to-noise spectra with wider wavelength coverage
and utilizing \ion{O}{2} and [\ion{O}{3}] emission lines.
This allows us to sample the free electrons with a 
wider energy coverage. For such an effort, very careful flux calibrations,
reddening corrections, and subtractions of stellar light contamination
are desirable.

\acknowledgments

The work described in this paper was substantially supported by
a grant from the Research Grants Council of the Hong Kong Special Administrative Region, China (project No. HKU 7073/11P). Part of the
financial support came from the HKU Small Project Funding
NO.201209176007 and the Key National Natural Science Foundation of China 
(No. 10933001).

\begin{figure*}
\epsfig{file=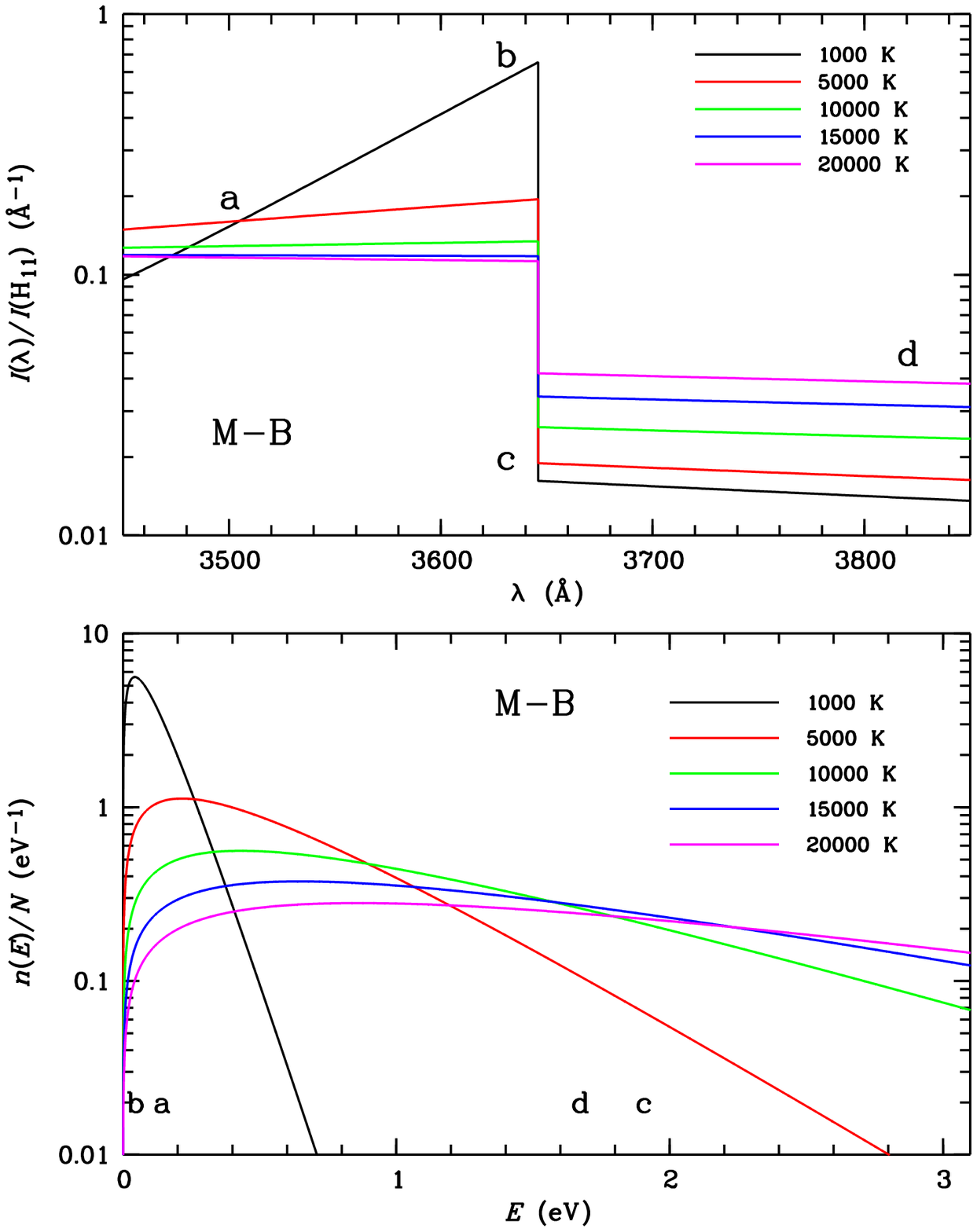,
height=19cm}
\caption{Theoretical hydrogen free-bound continua near the Balmer jump region
(upper panel) for single M-B electron distributions (lower panel) at different
temperatures. The spectral shapes in the upper panel can be characterized by 
the labeled `abcd', which trace the recombining electrons
with kinetic energies marked in the lower panel.
}
\label{MB}
\end{figure*}

\begin{figure*}
\epsfig{file=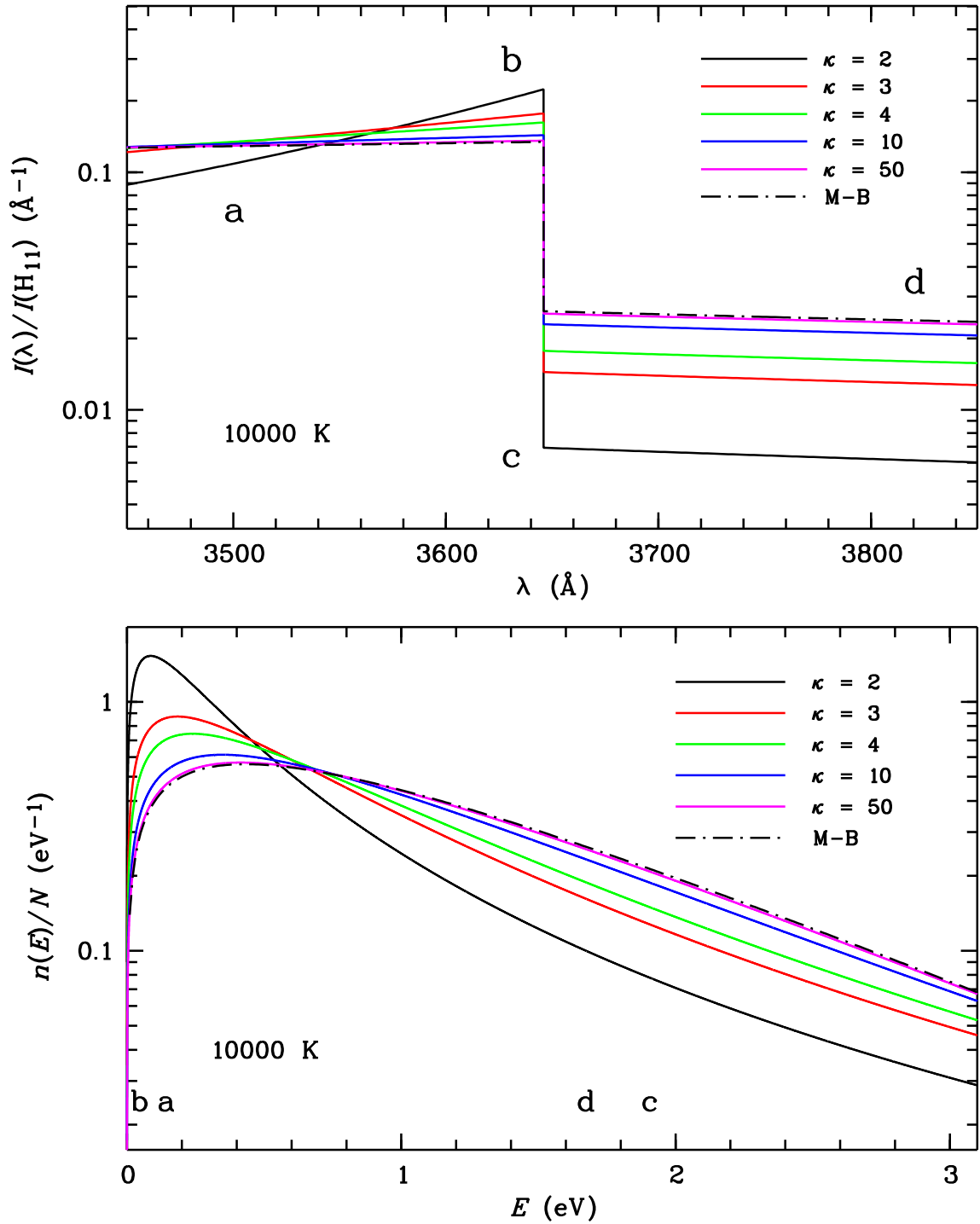,
height=19cm}
\caption{Theoretical hydrogen free-bound continua near the Balmer jump region
(upper panel) for $\kappa$ and M-B electron distributions (lower panel) at 
10000\,K but different $\kappa$ values. Other details are the same as in
Figure~\ref{MB}.
}
\label{kappa}
\end{figure*}

\begin{figure*}
\epsfig{file=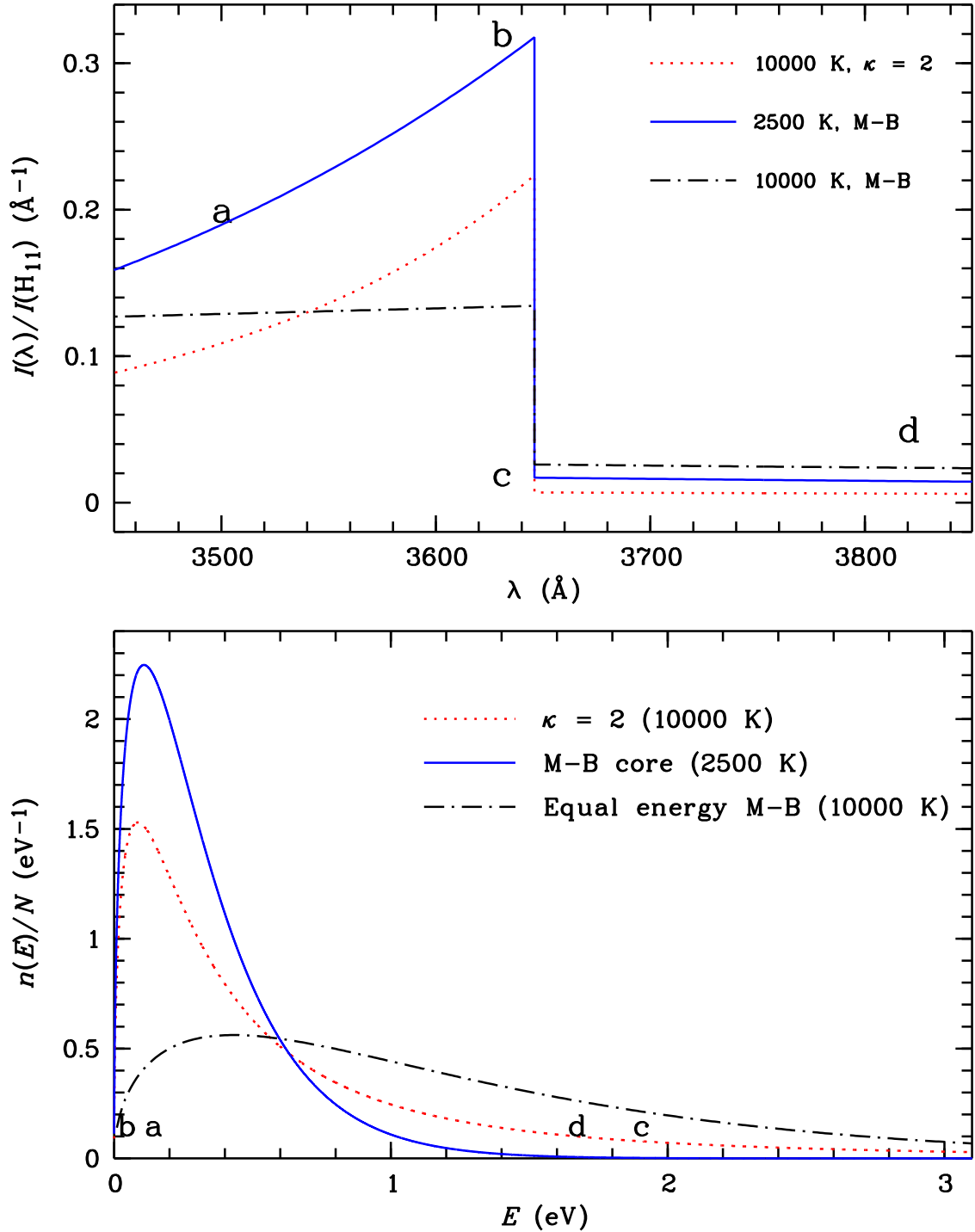,
height=19cm}
\caption{Theoretical hydrogen free-bound continua near the Balmer jump region
(upper panel) for $\kappa=2$ and M-B electron distributions at 10000\,K
as well as a `core' M-B electron distribution at 2500\,K (lower panel). 
In the lower panel, the areas under the three curves are all equal to 1,
and the M-B core can fit the low-energy region 
of the $\kappa=2$ distribution by scaling a factor of $<1$.
Other details are the same as in Figure~\ref{MB}.
}
\label{comp}
\end{figure*}

\begin{figure*}
\epsfig{file=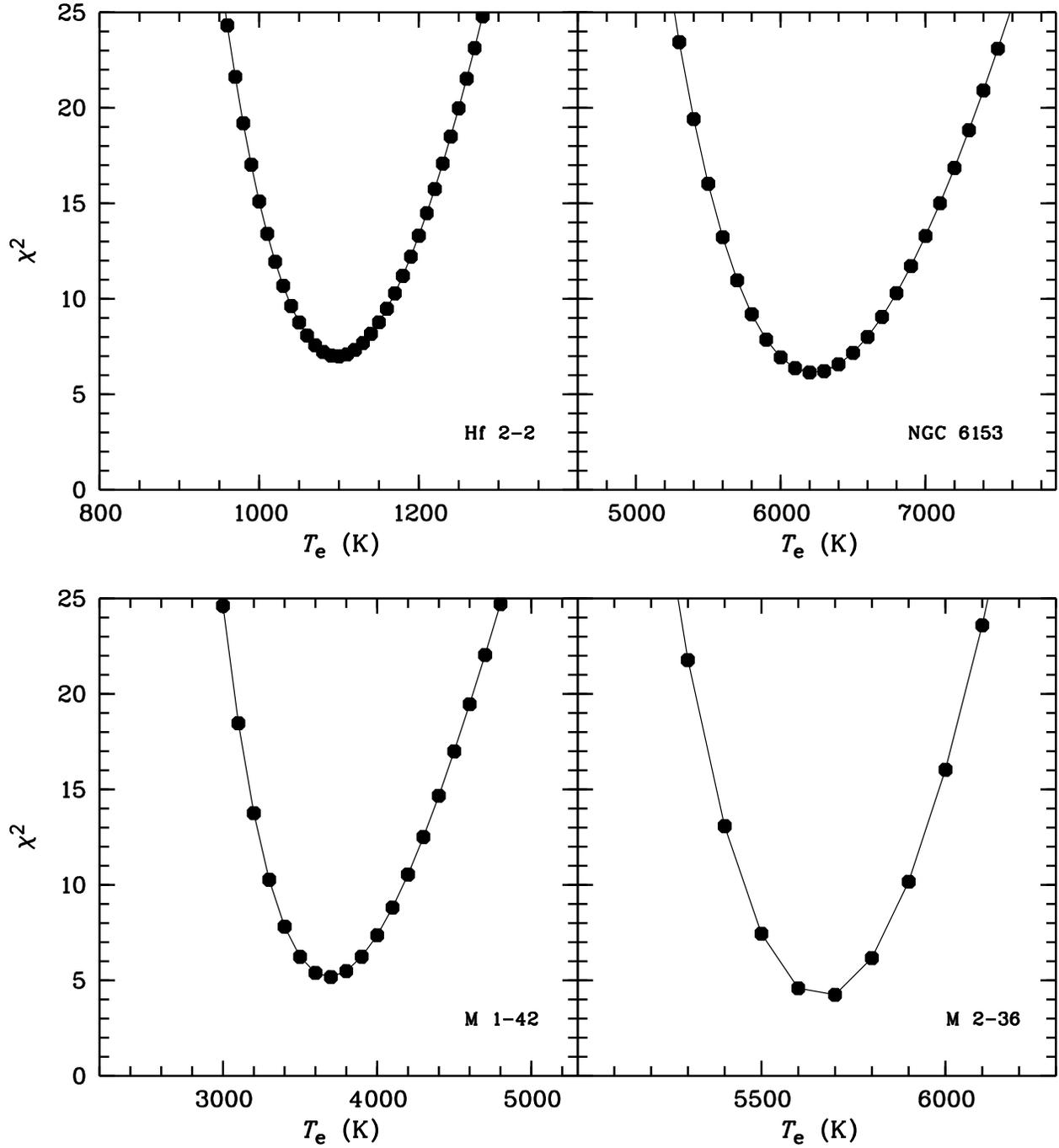,
height=18cm}
\caption{$\chi^2$ versus $T_{\rm e}$  for 
the fitting based upon single M-B electron distributions.
The filled circles represent an equal-interval scale of $T_{\rm e}$.
}
\label{one}
\end{figure*}

\begin{figure*}
\epsfig{file=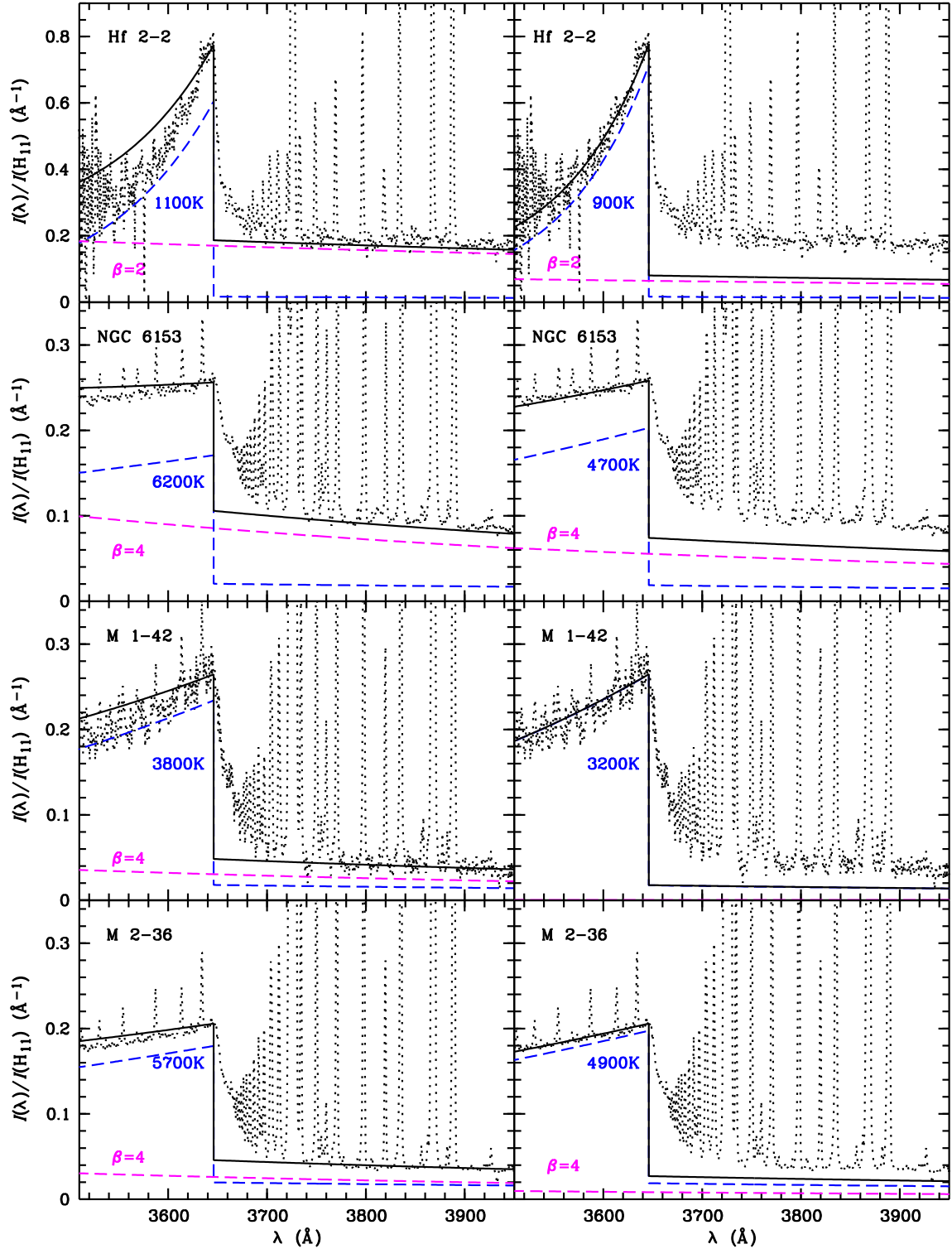,
height=20cm}
\caption{Fitting the Balmer jump ($J_B$; left panel) and the slope of
Balmer continua ($S_B$; right panel) using single M-B electron distributions.
The blue and magenta dashed curves denote the theoretical nebular continua and
scattered stellar light, respectively. Note that no scattered stellar light
is required to fit the $S_B$ value of M\,1-42.
The black solid and dotted curves are the synthetic and dereddened observed spectra, respectively.
Note that $J_B$ and $S_B$ (especially for Hf\,2-2) cannot be simultaneously 
fitted.
}
\label{fail}
\end{figure*}

\begin{figure*}
\epsfig{file=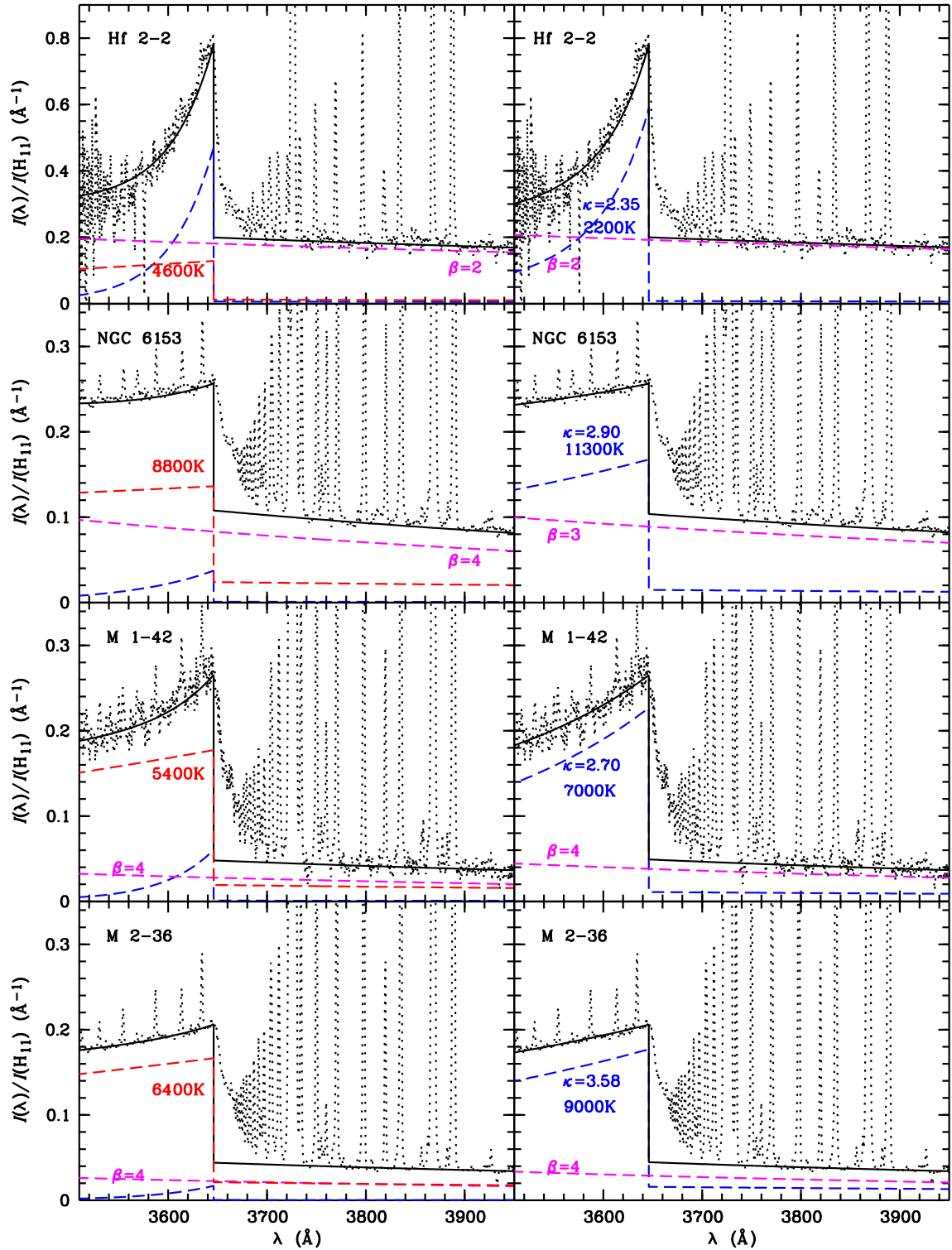,height=20cm}
\caption{Fitting the hydrogen continua near the Balmer jump region using
two-component models (left panel) and $\kappa$ electron distributions
(right panel). In the left panel, the blue and red dashed curves denote
the cold and hot nebular continua, respectively.
Other details are the same as in Figure~\ref{fail}.
}
\label{fitting}
\end{figure*}

\begin{figure*}
\begin{center}
\epsfig{file=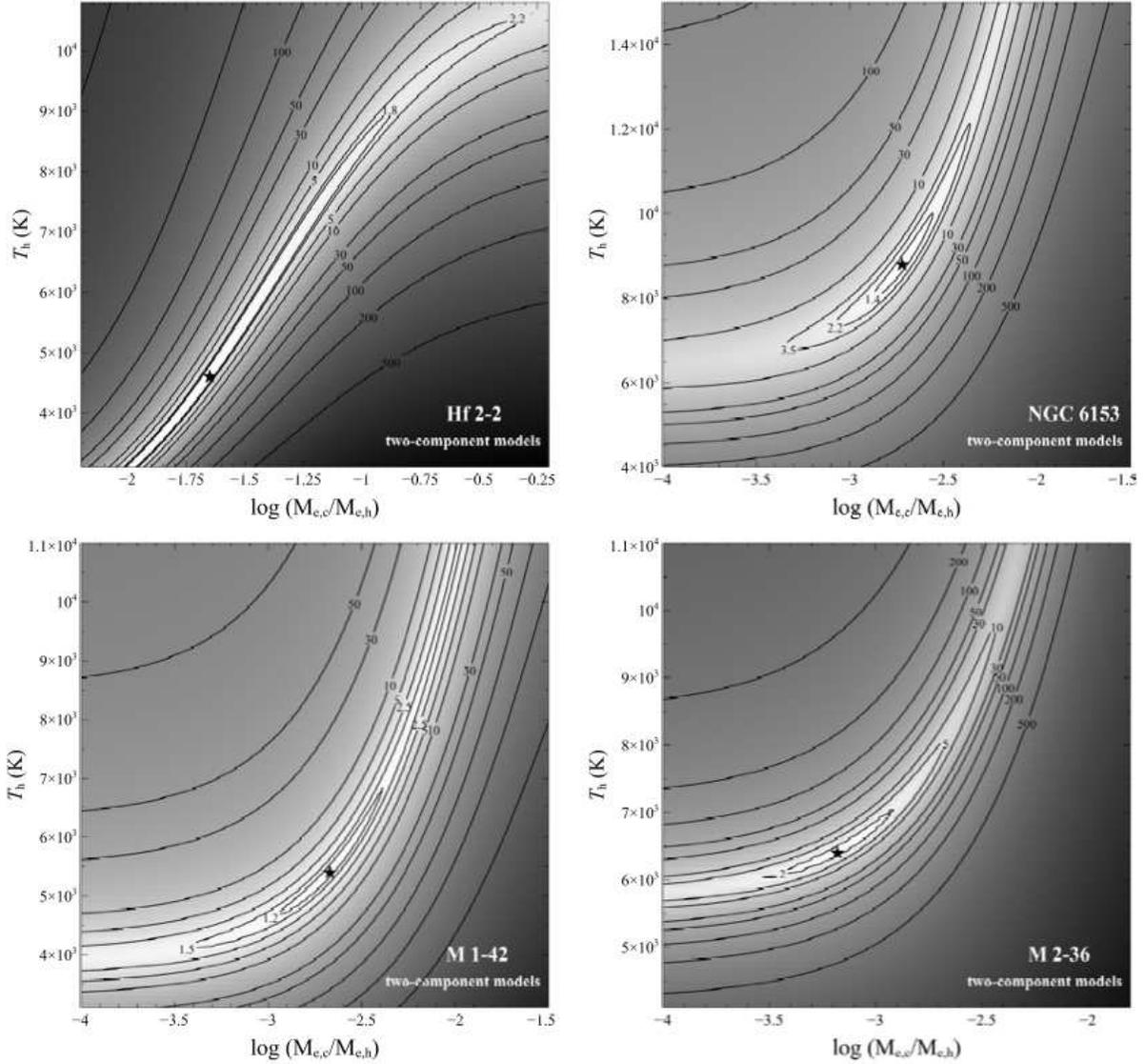,
height=15cm}
\end{center}
\caption{The $\chi^2$ contours of continuum fitting of
the two-component models as a function of $M_{e,c}/M_{e,h}$ and $T_h$.
The grey levels are proportional to the logarithm of $\chi^2$.
The stars mark the best-fit values, as shown in the left panels of Figure~\ref{fitting}.
\label{contourtwo}}
\end{figure*}

\begin{figure*}
\begin{center}
\epsfig{file=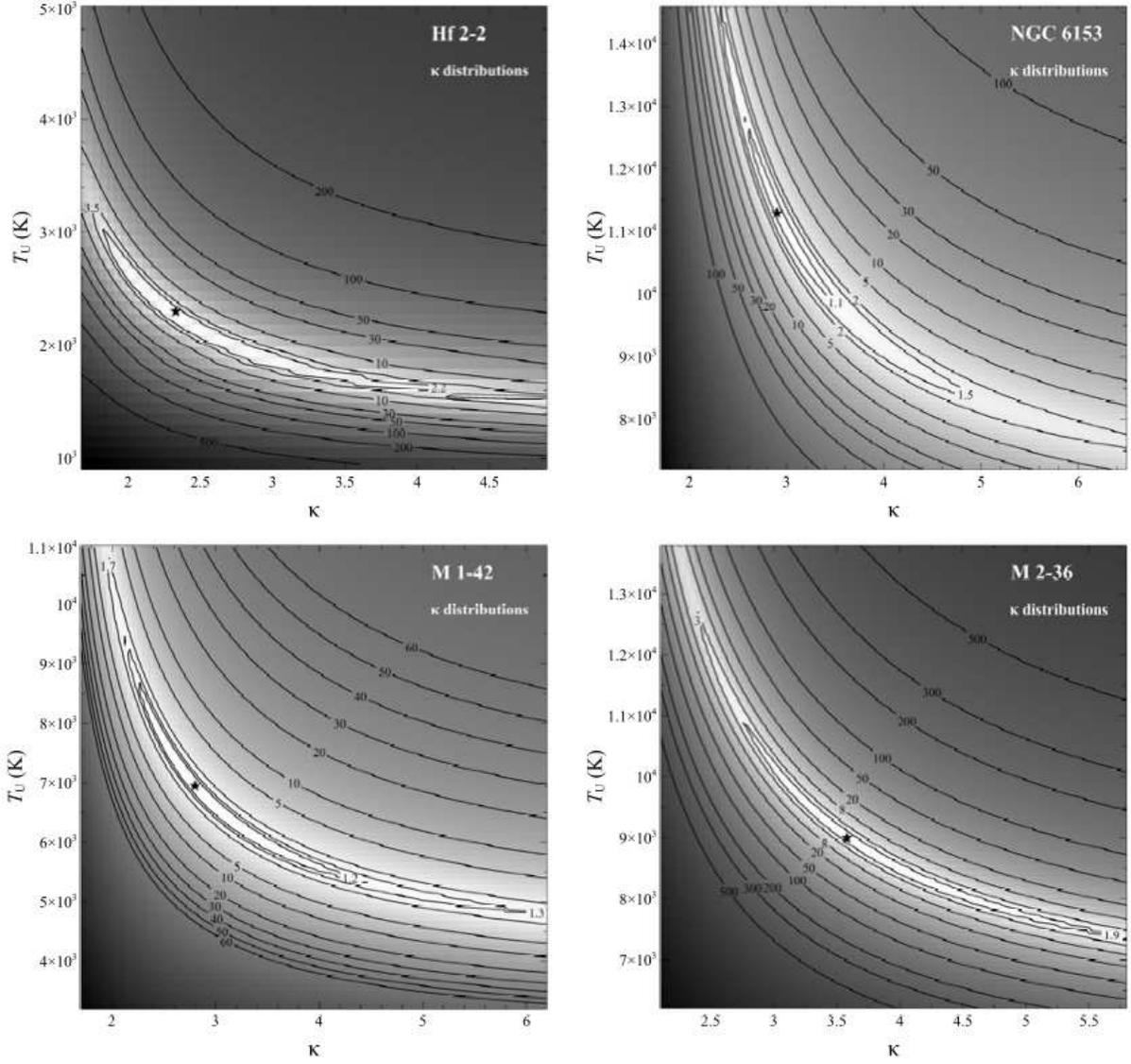,
height=15cm}
\end{center}
\caption{The $\chi^2$ contours of continuum fitting of the 
$\kappa$ electron energy distributions as a function of $\kappa$ and $T_U$.
The grey levels are proportional to the logarithm of $\chi^2$.
The stars mark the best-fit values, as shown in the right panels of Figure~\ref{fitting}.
\label{contourk}}
\end{figure*}

\begin{figure*}
\epsfig{file=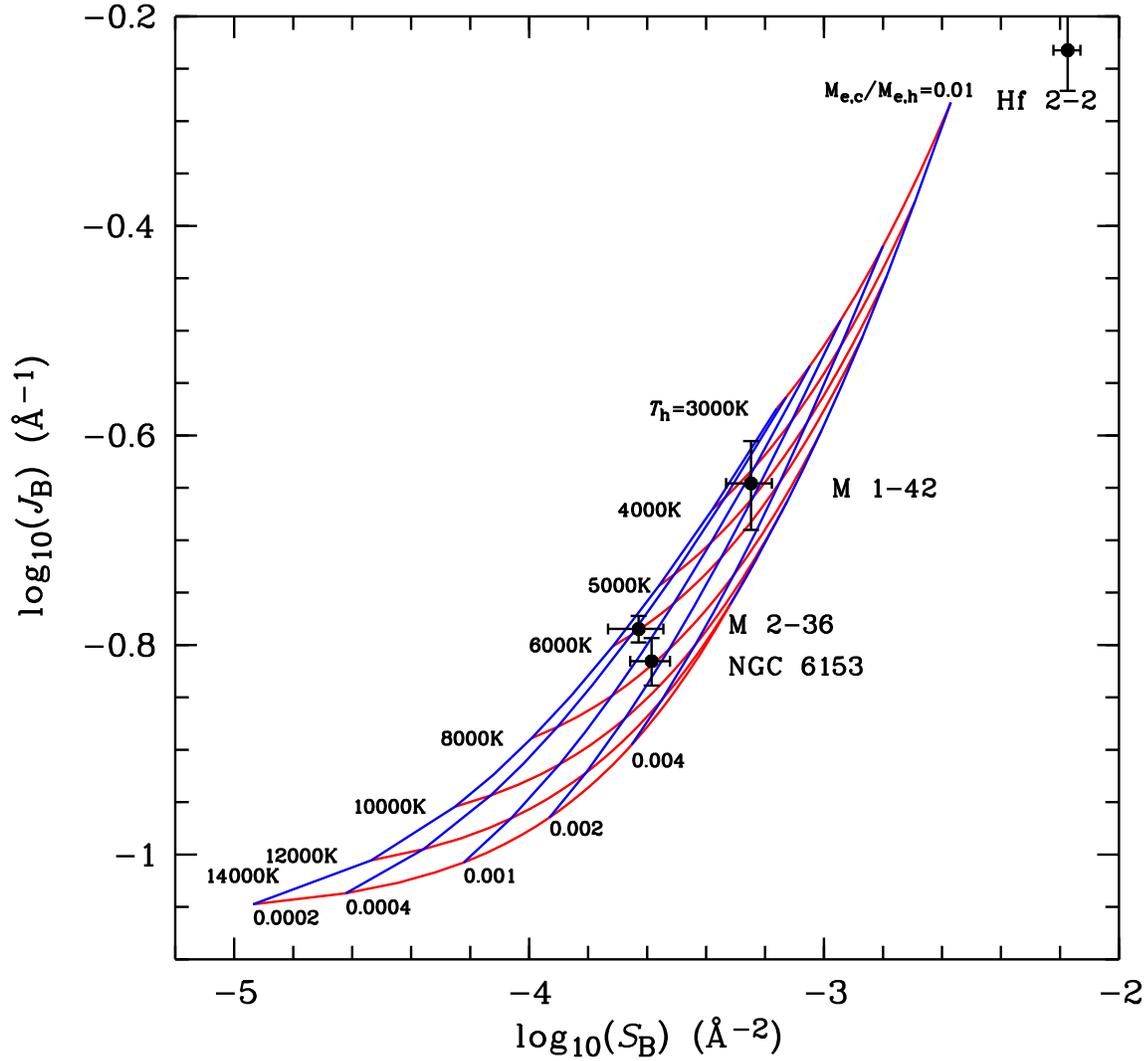,height=14cm}
\caption{$J_B$ versus $S_B$ as functions of M$_{e,c}$/M$_{e,h}$ values
from 0.0002--0.01 and $T_h$ from 3000--14000\,K in the two-component model. 
Note that the curve for M$_{e,c}$/M$_{e,h}=0.0002$  essentially coincides with 
that for single M-B electron distributions.  The measured values
of the four PNe are also
labeled.  Errors bars indicate the measurement uncertainties.
Note that because of the unsubtracted scattered stellar light,
the filled circles represent the lower limits of $S_B$.
}
\label{diabi}
\end{figure*}

\begin{figure*}
\epsfig{file=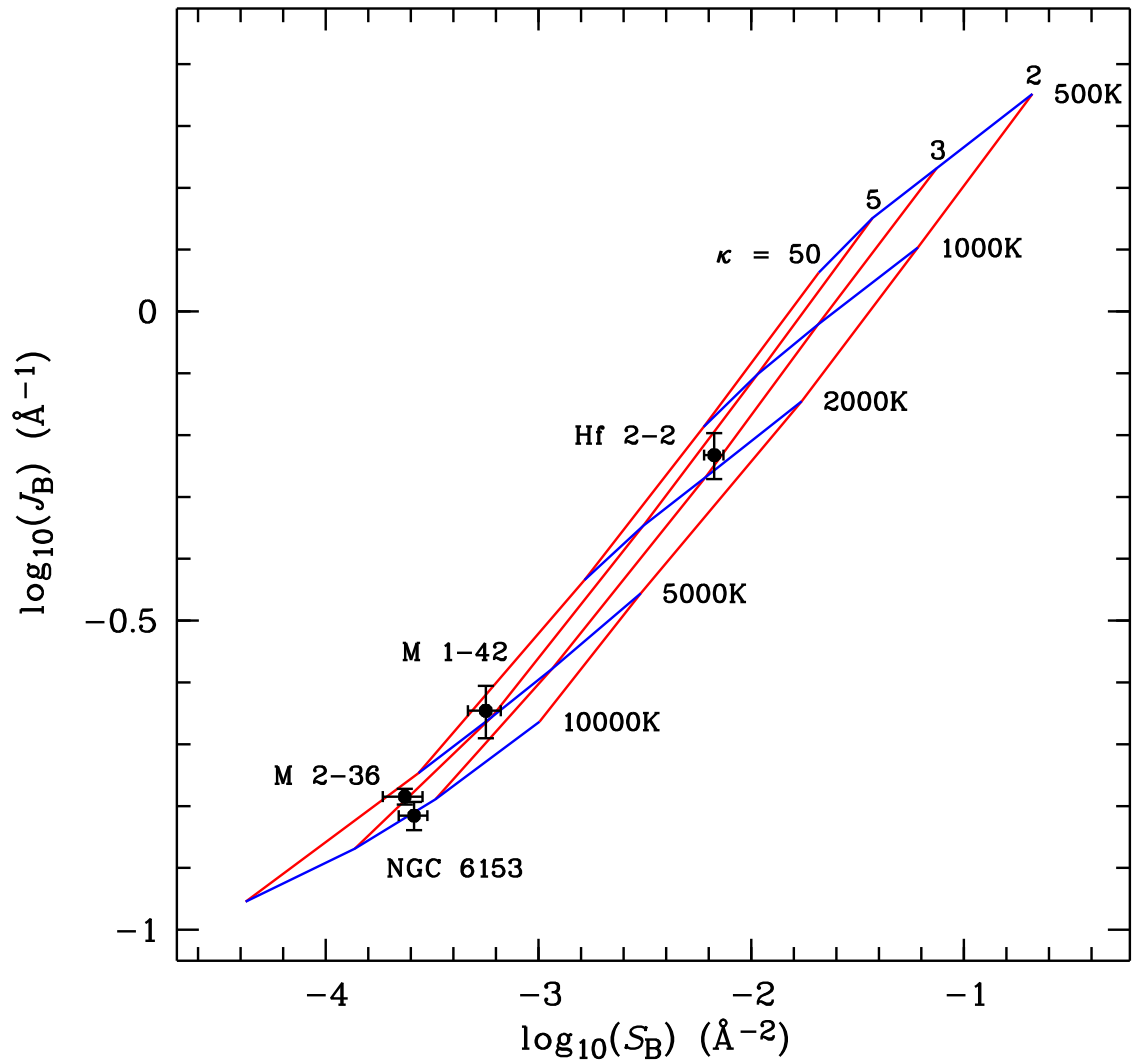,height=14cm}
\caption{$J_B$ versus $S_B$ as functions of $\kappa$ values
from 2--50 and kinetic temperatures from 500--10000\,K.
Note that the curve for $\kappa=50$ essentially coincides with that for 
single M-B electron distributions.
The description of measured values given for Figure~\ref{diabi} applies.}
\label{dia}
\end{figure*}

\begin{deluxetable}{lcccc}
\tablecaption{Fitting parameters
\label{para}}
\tablewidth{0pt}
\tablehead{
Parameters & Hf\,2-2 & NGC\,6153 & M\,1-42 & M\,2-36 \\
  }
\startdata
ADF$^a$ & 70 & 10 & 22 & 6.9\\
$T_{\rm e}$([\ion{O}{3}])$^a$ (K)& 8740 & 9110 & 9220 & 8380 \\
$T_{\rm e}$(BJ)$^a$ (K)& 930 & 6080& 3560& 5900\\
\vspace{0.005in}\\
 \multicolumn{5}{l}{Single M-B electron energy distributions}\\
$T_{\rm e}$ (K) &1100&6200&3800&5700\\
$\chi^2$  &6.99&6.14&5.17&4.24\\
\vspace{0.005in}\\
 \multicolumn{5}{l}{Two-component models}\\
$T_h$ (K) & 4600 & 8800 & 5400 &  6400 \\
$M_{e,c}/M_{e,h}$ ($10^{-2}$) & 2.24 & 0.20& 0.22 & 0.06 \\
$\chi^2$  &1.02&1.12&1.08&1.27\\
\vspace{0.005in}\\
 \multicolumn{5}{l}{$\kappa$ electron energy distributions}\\
$T_U$ (K) &2200&11300&7000&9000\\
$\kappa$  &2.35&2.90&2.70&3.58\\
$\chi^2$  &1.03&0.99&1.08&1.45\\
\enddata
\begin{description}
\item $^{a}$ From \citet{ls00,ll01,lb06}.
\end{description}
\end{deluxetable}

\end{document}